\documentclass[prb,twocolumn,showpacs,amssymb]{revtex4}

\begin{document}

\author{Baruch Horovitz{$^1$} and Pierre Le Doussal{$^2$} }
\affiliation{{$^1$} Department of Physics, Ben Gurion university,
Beer Sheva 84105 Israel} \affiliation{{$^2$}CNRS-Laboratoire de
Physique Th{\'e}orique de l'Ecole Normale Sup{\'e}rieure, 24 rue
Lhomond,75231 Cedex 05, Paris France.}

\title{Interference in presence of Dissipation}

\begin{abstract}
We study a particle on a ring in presence of various dissipative
environments. We develop and solve a variational scheme assuming
low frequency dominance. We analyze our solution within a
renormalization group (RG) scheme to all orders which reproduces a
2 loop RG for the Caldeira-Legget environment. In the latter case
the Aharonov-Bohm (AB) oscillation amplitude is exponential in
$-R^2$ where $R$ is the ring's radius. For either a charge or an
electric dipole coupled to a dirty metal we find that the metal
induces dissipation, however the AB amplitude is $\sim R^{-2}$ for
large $R$, as for free particles. Cold atoms with a large electric
dipole may show a crossover between these two behaviors.
\end{abstract}

\maketitle

The problem of interference and dephasing in presence of
dissipative environments is of significance for a variety of
experimental systems and a fundamental theoretical issue. The
experimental systems include mesoscopic rings embedded on various
surfaces where Aharonov-Bohm (AB) oscillations can be measured
\cite{web,jariwala}, and the related problem of decoherence at low
temperatures \cite{mohanty}. A different type of experimental
systems are cold atom traps created by atom chips
\cite{harber,jones,lin}, or trapped excited atoms with huge
electric dipoles \cite{hyafil}. The atom chip that produces a
magnetic or electric trap for the cold atoms necessarily also
produces noise. Our problem is then relevant for evaluating the
interference amplitude of cold atoms or molecules in presence of
such noise.

As an efficient tool for monitoring the effect of the environment
we follow a suggestion by Guinea \cite{guinea} to find the AB
oscillation amplitude as function of the radius R of the ring, as
measured by the curvature \cite{hofstetter,herrero} $1/B_c$ of the
ground state energy $E_0$ at external flux $\phi_x=0$, i.e.
$1/B_c=\partial^2E_0/\partial \phi_x^2|_0$. For free particles of
mass M this amplitude is the mean level spacing $\sim 1/MR^2$. Two
types of environments were suggested to lead to an anomalous
suppression, i.e. a stronger decrease of the oscillation amplitude
than $1/R^2$: system (i) is that of a Caledeira Legget bath and
system (ii) of a charge in a dirty metal environment.

System (i) is relevant to the Coulomb blockade problem
\cite{hofstetter,herrero,florens,buttiker} as well as to quantum
dots at a distance from metallic gates \cite{guinea2}. This
problem has been extensively investigated by instanton methods
\cite{panyukov,wang,beloborodov,lukyanov}, by RG methods
\cite{guinea,hofstetter,konig}, and by Monte Carlo (MC) methods
\cite{hofstetter,herrero,werner}. All methods show that $B_c$
increases exponentially with the dissipation strength $\alpha$,
i.e. $B_c\sim \alpha^{-\mu}e^{\pi^2\alpha}$, with differences in
the exponent $\mu$. In 2nd order renormalization group
\cite{hofstetter} (RG) $\mu=2$ , real time RG gives \cite{konig}
$\mu=6.5$, instanton methods give either \cite{panyukov} $\mu=2$
or \cite{wang,lukyanov} $\mu=3$, or \cite{beloborodov} $\mu=4$,
while MC suggests \cite{werner} $\mu=5$. This system was also
studied by a variational approach \cite{brown} which was solved
numerically showing a nonperturbative regime at strong $\alpha$.
Since $\alpha=\gamma R^2$ where $\gamma$ is a friction
coefficient, a length scale $\pi/\sqrt{\gamma}$ is identified
\cite{guinea}, beyond which the AB oscillations decay.

The system (ii) was investigated by RG methods \cite{guinea}
finding $B_c\sim R^{2+\mu '}$ with $\mu'\lesssim 1$ nonuniversal,
while MC data \cite{golubev} shows $\mu ' \approx 1.8$.
Furthermore, the MC data shows that at any finite temperature an
exponential form appears, with a temperature independent length.

In the present work we study also the effective action for an
electric dipole coupled to a dirty metal. This system (iii), which
is relevant to experiments on cold atoms \cite{hyafil} or
molecules, is found to induce dissipation on the dipole. We then
solve these systems by a variational method. We test our method on
system (i), which is extensively studied yet still controversial.
Our aim is, however, to develop an efficient method for a large
class of dissipative environments as in systems (ii) or (iii).

We show, within the variational method, that at zero temperature
the effective mass $B/R^2$ of the zero winding number sector
determines the curvature i.e. $B=B_c$. We find that the
variational method defines an RG scheme to all orders and it
reproduces the known RG equation \cite{hofstetter} to two loops in
system (i). In systems (ii) and (iii), we find that the
environment induces dissipation in the effective action, however
the effective mass remains $B/R^2\sim R^0$ for large $R$, as for
free particles. As a measurable result, we show that giant Rydberg
atoms \cite{hyafil} with huge dipole moments are sensitive probes
of metallic environments. They allow a crossover from system (i)
(at small $R$) with exponential decrease in its interference
amplitude to a large $R$ behavior with the much weaker $1/R^2$
behavior.

The time dependent angular position $\theta_m(\tau)$ of a particle
on the ring has in general a winding number $m$ so that
$\theta_m(\tau)=\theta(\tau)+2\pi m\tau/\beta$ where
$\theta(0)=\theta(\beta)$ has periodic boundary condition and
$\beta$ is the inverse temperature ($\beta\rightarrow\infty$
below). In presence of $\phi_x$ the partition sum has the form
\begin{eqnarray}\label{Z1}
Z&=&\sum_m e^{2\pi i m\phi_x-\frac{2\pi^2m^2MR^2}{\beta}
}Z_{m}\nonumber\\
Z_{m}&=&\int {\cal D}\theta
e^{-S_1\{\theta(\tau)\}-S_{int}\{\theta(\tau)+2\pi m\tau/\beta\}}
\end{eqnarray}
where in presence of a general dissipative bath the effective
action can be written in terms of Fourier coefficients
\cite{guinea} $\alpha_n$ \vspace{1cm}
\begin{eqnarray}\label{S}
&&S_1\{\theta_m\}=\int_0^{\beta}d\tau
\frac{1}{2}MR^2(\frac{\partial \theta_m}{\partial \tau})^2
\nonumber\\&& S_{int}\{\theta_m\} = \nonumber\\&& \sum_n
\alpha_n\int_0^{\beta}\int_0^{\beta}d\tau d\tau ' \frac{\pi ^2
\beta^{-2}\sin^2 \{\frac{n}{2}[\theta_m(\tau)-\theta_m(\tau
')]\}}{\sin ^2[\pi \beta^{-1}(\tau -\tau ')]}\nonumber\\&&
\end{eqnarray}
and $\alpha_n$ depend on the type of bath. At
$\tau\rightarrow\tau'$ (or at high frequencies $\omega$) one can
expand the $\sin^2(...)$ in (\ref{S}) and then $S_{int}\rightarrow
\sum_n\alpha_nn^2 \int d\omega |\omega||\theta_m (\omega)|^2$,
identifying a dissipative system. The Caldeira Legget bath has a
single $\alpha_n$
 with $\alpha_1=\gamma R^2$ while a charged
particle in a dirty metal bath has \cite{guinea,golubev}
$\alpha_n\approx \frac{2\gamma}{\pi r}\ln (r/n)$ for $1<n\lesssim
r$ and $\alpha_n\approx 0$
otherwise; here $\ell$ is the mean free path, $k_F$ is the Fermi
wavevector, $r=R/\ell$ and $\gamma=3/(8k_F^2\ell^2)$.

As a new realization of Eq. (\ref{S}) we consider an electric
dipole $p$ perpendicular to the plane of the ring and coupled to a
dirty metal. The interaction with the fluctuating electric field
${\bf E}({\bf r},\tau)$ is $p\int_0^{\beta}E_z({\bf
R}(\tau),\tau)d\tau$ where ${\bf R}(\tau)=R[\cos \theta(\tau),\sin
\theta(\tau)]$ is the particle's position on the ring. After a
Gaussian average we have $S_{int}=\frac{1}{2}p^2\int\int d \tau
d\tau' f({\bf X},\tau-\tau')$ where ${\bf X}={\bf R}(\tau)-{\bf
R}(\tau')$, and
\begin{eqnarray}\label{dipole}
f({\bf X},\tau)&=&\langle E_z({\bf R}(\tau),\tau) E_z({\bf
R}(0),0)\rangle\nonumber\\
&=&\sum_{q,\omega_n}e^{i{\bf q}\cdot {\bf
X}-i\omega_n\tau}\frac{4\pi q_z^2}{q^2\epsilon(i|\omega_n|,q)}
\end{eqnarray}
with $\epsilon(i|\omega_n|,q)$ being the dielectric function
\cite{golubev}, $\omega_n=2\pi n/\beta$. The frequency sum yields
the form (\ref{S}) with
\begin{equation}\label{dipole1}
\frac{3}{8k_F^2\ell^2}\frac{p^2}{e^2\ell^2}
\{1-(4r^2\sin^2\frac{z}{2}+1)^{-3/2}\}=\sum_n\alpha_n\sin^2n\frac{z}{2}\,.
\end{equation}
Hence, for large $r$, $\alpha_n\sim\frac{1}{r}(1-\frac{n^2}{r^2})$
for $n<r$ and $\alpha_n\approx 0$ otherwise.

The variational method \cite{feynman} for $Z_m$ finds the best
Gaussian approximation, i.e.
$S_0=\frac{1}{2\beta}\sum_{\omega_n}G^{-1}(\omega_n)|\theta(\omega_n)|^2$
so that the variational free energy $\beta F_{var}=\beta
F_0+\langle S_1+S_{int}-S_0\rangle_0$ is minimized; here $\langle
... \rangle_0$ is an average with respect to $\exp(-S_0)$ and
$F_0$ is the free energy corresponding to $S_0$. The method is
tested below by RG results, where available, and is found to
reproduce these RG results. The interaction term is then
\begin{widetext}
\begin{equation}\label{Sint}
\langle S_{int}\rangle_0=\beta \sum_n\alpha_n \int_{0}^{\beta}
\frac{d\tau}{2\tau^2}\{1-\cos (2\pi
nm\tau/\beta)e^{-n^2\int_{\omega}G(\omega)[1-\cos (\omega\tau)]}\}
\end{equation}
where $\int_{\omega}=\int d\omega/2\pi$ and the variational
equation $\delta F_{var}/\delta G(\omega_n)=0$ becomes
\begin{equation}\label{G}
G^{-1}(\omega)=MR^2\omega^2+2\sum_n\alpha_n
n^2\int_{1/\omega_c}^{\infty}d\tau
\frac{1-cos(\omega\tau)}{\tau^2}\cos (2\pi
nm\tau/\beta)e^{-n^2\int_{\omega_1}G(\omega_1)[1-\cos
(\omega_1\tau)]}\,.
\end{equation}
\end{widetext}
Here the limit $\beta\rightarrow \infty$ is taken (except for the
$m$ dependent term) and a cutoff $\omega_c$ is introduced to
control the short time behavior. This cutoff represents a high
frequency limit of the bath degrees of freedom.

In the following we will study the variational equation with
$m=0$. To justify this, we show now that the effective mass $B$ of
the $m=0$ system is indeed what is needed to find the AB
oscillation amplitude at $\beta\rightarrow \infty$. The effective
mass is defined by $G^{-1}(\omega)=B\omega^2$ in the limit
$\omega\rightarrow 0$ and is identified from Eq. (\ref{G}) at
$\beta\rightarrow \infty$ as
\begin{equation}\label{B0}
B=MR^2+\sum_n\alpha_nn^2\int_0^{\infty} d\tau
e^{-n^2\int(d\omega/2\pi)G(\omega)[1-\cos (\omega\tau)]}
\end{equation}
We use here $\langle m^2 \rangle \sim \beta$ (Eq. \ref{Z1}) and
convergence in $\tau$ due to the exponent in (\ref{G}) $\sim
\exp(-n^2\tau/B)$. Hence $\cos (2\pi nm\tau/\beta)\rightarrow
1+O(1/\beta)$ in Eq. (\ref{G}) and the effective mass $B$ is $m$
independent.

The AB oscillation amplitude is usually measured by by the
curvature $1/B_c=-\partial^2E_0/\partial^2\phi_x|_0=4\pi^2\langle
m^2 \rangle /\beta |_0$. To identify $B_c$ we expand $\ln
Z_m=-\beta F_{var}(m)$ in $m/\beta$ and since at $m=0$ we have
$\partial G(\omega)/\partial m=0$ and $\partial
F_{var}(m)/\partial G=0$ (the variational condition) the leading
term is from expanding Eq. (\ref{Sint})
\begin{equation}
\beta F_{var}(m)=\beta F_{var}(0)+
\frac{4\pi^2}{\beta}(B-MR^2)m^2+O(m^4/\beta^3)\,.
\end{equation}
The effect of $S_{int}$ in the partition sum Eq. (\ref{Z1}) is
therefore to replace the factor $2\pi^2MR^2m^2/\beta$ by
$2\pi^2Bm^2/\beta$, i.e. the response to an external flux is that
of a free particle with a mass renormalized to $B/R^2$. Our task
is therefore to study the $m=0$ system and find this renormalized
mass.

We note that at finite $\beta<B/n^2$ the $\tau$ integrals are not
suppressed by the exponential factor and $\langle
S_{int}\rangle_0\approx \pi^2 \sum_n\alpha_n n |m|$, corresponding
to instanton trajectories \cite{panyukov,wang,beloborodov}. Hence
for large $\alpha_n$ only the lowest $m$ sectors contribute, in
contrast with low temperatures where $\langle m^2\rangle\sim
\beta/B\rightarrow \infty$; This defines a crossover temperature
$\beta^*\approx B/n^2$

We assume now that the frequency integral in Eq. (\ref{G}) is
dominated by low frequencies and derive a simplified variational
equation. Consider first the regime $\omega<\omega_c$, but
$\omega$ is not too small, i.e. $\ln (\omega_c/\omega) \approx 1$.
This is a perturbative regime where the significant low
frequencies are not yet manifested. For large dissipation
coefficients $\alpha_n$ the term in the exponent of Eq. (\ref{G})
is small, and $G^{-1}(\omega)=MR^2\omega^2+\pi \omega
\sum_n\alpha_n n^2$. This form shows that the $\omega^2$ term acts
as cutoff $\omega_c'$ in $\int_{\omega} G(\omega)$ with
$\omega_c'=\pi  \sum_n\alpha_n n^2/(MR^2)$; we choose $\omega_c$
as the lowest of the original $\omega_c$ and $\omega_c'$ and
ignore the bare $\omega^2$ term. The next order in perturbation is
then
\begin{equation}\label{Gpert2}
G^{-1}(\omega)=\pi \omega \sum_n\alpha_n
n^2[1-\frac{n^2}{\pi^2\sum_m\alpha_m m^2}\ln
\frac{\omega_c}{\omega}]
\end{equation}
which identifies the perturbation parameter, i.e. the perturbative
$\omega$ range is large for system (i), while for systems (ii) and
(iii) the $n\approx r$ terms require $\ln \omega_c/\omega\ll 1$.

We proceed now to the significant range of $\omega\ll \omega_c$
and assume the general form
\begin{eqnarray}\label{forms}
G^{-1}(\omega)&=&f(\omega) \qquad   \omega_0<\omega\ll \omega_c
\nonumber\\
G^{-1}(\omega)&=& B\omega^2    \qquad  \omega<\omega_0 \,.
\end{eqnarray}
It is convenient, for $\omega>\omega_0$, to study a derivative of
Eq. (\ref{G}) for which the oscillating $\sin (\omega \tau)/\tau$
 is replaced by $1$ while the $\tau$ integration
acquires a cutoff $\tau<1/\eta_1 \omega$, such that the result
should not be sensitive to $\eta_1\approx 1$.

We assume now that the the integral in the exponent of (\ref{G})
is dominated by $\omega_1>1/\tau$ where the $\cos \omega_1\tau$
averages to zero. For $\omega\gtrsim\omega_0$ this requires
$1/B\omega_0\ll \int_{\omega_0}^{\omega_c}d\omega_1/f(\omega_1)$
[condition (i)]. Introducing a second cutoff uncertainty $\eta_2$
we obtain in terms of $\omega_2=1/\tau$
\begin{equation}\label{f2}
f'(\omega)=2\omega\sum_n\alpha_n
n^2\int_{\eta_1\omega}^{\omega_c}\frac{d\omega_2}{\omega_2^2}
e^{-n^2\int_{\eta_2\omega_2}^{\omega_c}d\omega_1/\pi
f(\omega_1)}\,.
\end{equation}
Taking $d/d\omega$ we obtain our main equation for $f(\omega)$,
\begin{equation}\label{f4}
f'(\omega)=\pi \eta \sum_n\alpha_n n^2
e^{-n^2\int_{\omega}^{\omega_c}d\omega_1/\pi f(\omega_1)}
\end{equation}
where $\omega f''(\omega)\ll f'(\omega)$ is assumed [condition
(ii)]. Here
 $\eta_1=2/(\pi \eta)$ and $\eta_1\eta_2=1$ are chosen, to
connect smoothly with the perturbative regime where
$f'(\omega_c)=\pi \eta\sum_n\alpha_n n^2$; from Eq. (\ref{Gpert2})
\begin{equation}\label{eta}
\eta=1+\frac{\sum_n\alpha_n n^4}{(\pi\sum_n\alpha_n n^2)^2} \,.
\end{equation}

A similar analysis for the range $\omega<\omega_0$ leads to
$f'(\omega_0)=\eta' B\omega_0$ [note the similarity of (\ref{B0})
and (\ref{f4})] with $\eta'$ reflecting cutoff uncertainties.
Hence Eq. (\ref{f4}) is to be solved with the boundary conditions
[$\eta$ given by (\ref{eta})]
\begin{eqnarray}\label{allbc}
f(\omega_c)&=&\pi\omega_c\sum_n\alpha_n n^2\,\,\,;\qquad f(\omega_0)=B\omega_0^2 \nonumber\\
f'(\omega_c)&=&\eta\pi\sum_n\alpha_n n^2\,\,\,;\qquad
f'(\omega_0)=\eta' B\omega_0 \,.
\end{eqnarray}

We show now that the variational Eq. (\ref{f4}) can be solved by
an RG process. The latter identifies a change in the cutoff
$d\omega_c=\omega_c'-\omega_c$ combined with a change in the
couplings $d\alpha_n=\alpha_n'-\alpha_n$ such that Eq. (\ref{f4})
for $f'(\omega)$ is unchanged. In terms of $d\ell =-d(\ln
\omega_c)$ this yields an RG equation to all orders (provided
$\eta(\alpha_m)$ is known)
\begin{equation}\label{RG}
\frac{d\alpha_n}{d\ell}+\frac{\alpha_n}{\eta}\sum_m\frac{
\partial \eta}{\partial
\alpha_m}\frac{d\alpha_m}{d\ell}=-\frac{n^2\alpha_n}{\pi^2\sum_m\alpha_m
m^2}\,.
\end{equation}
which to lowest order ($\eta=1$) agrees with the RG proposed by
Guinea \cite{guinea}. In system (i) with a single $\alpha$ we
obtain to order $1/\alpha$ (and $\alpha\equiv \alpha_1$ for
brevity)
\begin{equation}\label{RG1}
\frac{d\alpha}{d\ell}=-\frac{1}{\pi^2}-\frac{1}{\pi^4\alpha}
\end{equation}
which amazingly is precisely the 2 loop RG result
\cite{hofstetter} (requiring 14 diagrams). For system (ii)
$\sum_n\alpha_n n^4/(\sum_n\alpha_n n^2)^2=O(1)$ is independent of
the large parameter $r$. Hence there is no expansion parameter for
the RG, yet the variational method is useful as shown below.

We proceed to solve the nontrivial Eq. (\ref{f4}) for the
Caldeira-Legget system. Differentiating Eq. (\ref{f4}) we obtain
$f''(\omega)=\frac{f'(\omega)}{\pi f(\omega)}$ which upon
integration yields
\begin{equation}\label{s2}
f'(\omega)=\pi^{-1}\ln [Kf(\omega)]
\end{equation}
where from the boundary conditions (\ref{allbc})
\begin{equation}\label{s8}
K=\frac{e^{\pi^2\alpha\eta}}{\pi\alpha\omega_c} \,.
\end{equation}

A solution of (\ref{s2}) requires an asymptotic expansion of the
Log integral $\int df/\ln f$, an expansion that is not available
in standard textbooks. We develop here a large $\alpha$ expansion
using the following idea: In terms of $f(\omega)=\omega
g(K\omega)$ we have
\begin{equation}\label{g}
g(x)+xg'(x)=\pi^{-1}\ln [xg(x)]
\end{equation}
The boundary condition at $\omega=\omega_c$ can be written as
\begin{equation}\label{gb}
g(\frac{e^{\eta\pi^2\alpha}}{\pi\alpha})=\pi\alpha \,.
\end{equation}

 We claim that if the function $g(K\omega)$ is  chosen such that it does not
depend explicitly on $\alpha$, except through its argument
$K(\alpha)$, then a useful large $\alpha$ expansion is generated.
The boundary condition (\ref{gb}) becomes a functional relation
involving $\eta(\alpha)$. To show our claim we use the boundary
condition (\ref{gb}) $g(x_c)=\pi\alpha$ at $x_c=K\omega_c$, its
derivative
 $x_cg'(x_c)=\pi\frac{K(\alpha)}{K'(\alpha)}$, and
  $f'(\omega_c)=\pi\eta\alpha=g(x_c)+x_cg'(x_c)$ to yield
\begin{equation}\label{g1}
\eta=1+\frac{K(\alpha)}{\alpha K'(\alpha)}=1+\frac{1}{\pi^2
\alpha\eta -1+\pi^2\alpha^2\frac{d\eta}{d\alpha}} \,.
\end{equation}
This relation generates a large $\alpha$ expansion with the
leading form $\eta=1+(\pi^2\alpha)^{-1}+O(\alpha)^{-2}$,
consistent with the perturbation expansion Eq. (\ref{eta}). It is
remarkable that the initial values for (\ref{s2}) as given by the
perturbation expansion are precisely such as to allow for an
asymptotic expansion of (\ref{g1}).

We note that combining the RG equation (\ref{RG}) with (\ref{g1})
leads to yet another remarkable relation
\begin{equation}\label{RG2}
\frac{d\alpha}{d\ell}=\alpha[1-\eta(\alpha)]
\end{equation}
where $\eta(\alpha)$ solves (\ref{g1}). Hence $1-\eta(\alpha)$ is
the exact $\beta$ function within the variational method. In view
of its success in reproducing the 2 loop result \cite{hofstetter}
it may hold even to higher orders in RG for the original action
(\ref{S}).

For $\omega<\omega_c$ we choose ${\bar \alpha}(\omega)$ such that
$K[{\bar \alpha}(\omega)]\omega_c=K(\alpha)\omega$. The boundary
condition (\ref{gb}) with ${\bar \alpha}$ produces then the
solution $f(\omega)=\omega g(K\omega)=\pi\omega {\bar
\alpha}(\omega)$ with ${\bar \alpha}(\omega)$ the solution of
\begin{equation}\label{alphabar}
K\omega=\frac{e^{\pi^2{\bar \alpha}(\omega)\eta[{\bar
\alpha}(\omega)]}}{\pi {\bar \alpha}(\omega)} \,.
\end{equation}
Inverting this relation we find
\begin{eqnarray}\label{g2}
f(\omega)&=&\pi\omega{\bar
\alpha}(\omega)=\frac{\omega}{\pi\eta}\ln (\pi{\bar \alpha}
K\omega)=
 \frac{\omega}{\pi\eta}\ln
(\frac{K\omega}{\eta\pi}\ln (\pi{\bar \alpha}
K\omega))\nonumber\\&=&\frac{\omega}{\pi\eta}\ln
(\frac{K\omega}{\eta\pi}\ln (\frac{K\omega}{\eta\pi}...))
\end{eqnarray}
and at least two $\ln$ embeddings are needed for a large $\alpha$
solution.

To identify the effective mass $B$ we note that the boundary
condition for (\ref{s2}) at $\omega_0$ is $K\omega_0=e^{\eta'\pi
B\omega_0}/B\omega_0$ so that $g(K\omega_0)=B\omega_0$ becomes
$g(e^{\eta'\pi B\omega_0}/B\omega_0)=B\omega_0$. This equation
does not involve the large parameter $\alpha$, hence
 $B\omega_0\approx 1$, $K\omega_0\approx 1$, and the effective mass is
\begin{equation}\label{B1}
B\approx \frac{1}{\omega_0}\approx
\frac{e^{\pi^2\alpha}}{\alpha\omega_c}\,.
\end{equation}

Condition (i) for (\ref{f4}) is satisfied if $\ln \alpha_1 \gg 1$,
while condition (ii) requires $\ln \omega/\omega_0 \gg 1$; this is
valid in the exponentially large range $\alpha
\omega_0<\omega<\omega_c$ while in the relatively small range
$[\omega_0,\alpha\omega_0]$ $O(1)$ changes occur in Eq.
(\ref{f4}).

We conclude that the effective mass is exponentially large, Eq.
(\ref{B1}), and that the power of the prefactor is $\mu=2$, as in
the 2nd order RG result \cite{hofstetter}; (We use here the cutoff
$\omega_c=\omega_c'=\pi\alpha /MR^2$ adding $1$ to the power of
$\alpha$ in (\ref{B1})). The crossover temperature is
$1/\beta^*\approx \omega_0$ where $\langle m^2 \rangle $ drops
from the divergent $\sim \beta/B$ to a small $\sim 1/\alpha$ value
at high temperatures.

We proceed now to multi $\alpha_n$ problems. A generalized
asymptotic expansion in the parameter $(\sum_n \alpha_n
n^2)^2/\sum_n\alpha_n n^4$ is possible. Replacing
$\alpha_n\rightarrow\gamma\alpha_n$ we obtain that Eq. (\ref{f4})
is equivalent to
\begin{equation}\label{eta2}
\eta=1+\frac{1}{\gamma\pi^2(\gamma\frac{d\eta}{d\gamma}+\eta)\frac{(\sum_n\alpha_n
n^2)^2}{\sum_n\alpha_n n^4}-1}\,.
\end{equation}
It is again remarkable that the perturbation expansion (\ref{RG2})
allows an asymptotic expansion of (\ref{eta2}) for large $\gamma$,
i.e. from (\ref{eta}) $\gamma^2d\eta/d\gamma\sim O(\gamma^0)$
produces higher order terms.

We show now that for the charge or dipole in a dirty metal
environment, systems (ii) and (iii), the dependence on the radius
is $B\sim R^2$ as for the free particle. For both cases and for
$r\gg 1$ $\alpha_n=r^{-1}\alpha^*(n/r)$ with $\alpha^*(n/r)$
decaying to $0$ after a large number of terms $n\approx r$. Hence
the action in Eq. (\ref{S}) has the form, for the $m=0$ winding
number,
\begin{eqnarray}
S_{int}\{\theta_0\}&=&\sum_{n=1}^{n=r}(1/r)\alpha^*(n/r)\bar{S}\{
n\theta_0(\tau)\}\nonumber\\&& \rightarrow \int_0^1 dx \alpha^*(x)
\bar{S}\{x\bar {\theta}_0(\tau)\}
\end{eqnarray}
 where $\theta(\tau)$ is rescaled, ${\bar \theta}_0(\tau)=r\theta_0(\tau)$. The action (including
the free term $S_1$) is then r independent and therefore the
effective mass for $[{\bar \theta}_0(\tau)]^2$ is r independent,
which after unscaling yields $B\sim r^2$. We rely here on the
variational scheme only to the extent that it shows that $B_c$ can
be deduced from the $m=0$ sector, i.e. $B=B_c$. For an actual
solution for $f(\omega)$ we can imagine starting from a large
$\gamma$ and integrating (\ref{eta2}) to an actual value of
$\gamma\lesssim 1$. Since (\ref{eta2}) is $r$ independent for
large $r$, the resulting $\eta$ will also be $r$ independent.

We note that for $r\lesssim 1$ the dipole problem reduces to that
of the Caldeira Legget system (i) with $\alpha =
\frac{9r^2}{4k_F^2\ell^2}(\frac{p}{e\ell})^2$. Hence for large
$p$, as for the giant Rydberg atoms \cite{hyafil}, one can be in
the regime of large $\alpha$ showing the exponential dependence of
(\ref{B1}). Upon further increase of $r$ a crossover to the free
particle form is predicted with $B\sim r^2$. We also note that the
instanton crossover temperature $\beta^*=B/r^2$ is $r$
independent. At temperatures above $1/\beta^*$ we have $\langle
m^2 \rangle \sim e^{-\gamma' r}$ with
$\gamma'=\frac{3\pi}{2k_F^2\ell^2}$ for a charged particle, as in
Ref. [\onlinecite{golubev}], while $\gamma'=\frac{3\pi
}{k_F^2\ell^2}(\frac{p}{e\ell})^2$ for a charge dipole.

In conclusion, we developed and solved a variational scheme for a
large class of dissipative systems. The solution provides an RG of
a high order, provided that a coefficient $\eta$ is derived to
high order in the perturbative regime. We also found an efficient
asymptotic expansion for the relevant type of differential
equations. We applied our method to the Caldeira-Legget system and
found that its AB amplitude behaves as $\sim R^2 e^{-\pi^2\gamma
R^2}$. For a charged particle or a charged dipole in a dirty metal
environment we find for large $R$ an AB amplitude of $\sim R^{-2}$
as for free particles, in contrast with previous results
\cite{guinea,golubev}. Huge charge dipoles \cite{hyafil} in a
dirty metal environment can show a variety of behaviors and
provide therefore a valuable probe of dissipative environments.


\begin{acknowledgments}
We thank for valuable discussions with D. Cohen, F. Guinea, A. D.
Zaikin, G. Zar\'{a}nd, J. Dalibard and G. Nogues. This research
was supported by THE ISRAEL SCIENCE FOUNDATION founded by the
Israel Academy of Sciences and Humanities, the DIP German Israeli
program and by the French ANR.
\end{acknowledgments}

\end{document}